# Lead germanate: an advanced material for infrared electro-optic modulators

Nataliya Demyanyshyn
Karpenko Physico-Mechanical Institute
Lviv, Ukraine
nmdem@ipm.lviv.ua

Oleh Buryy
Lviv Polytechnic National University
Lviv, Ukraine

Bohdan Mytsyk
Karpenko Physico-Mechanical Institute
Lviv, Ukraine

Anatoliy Andrushchak
Lviv Polytechnic National University
Lviv, Ukraine

*Abstract* – **the optimal conditions of electro-optic interaction in lead germanate crystals are determined and the geometries of the electro-optic cells of modulators of infrared electromagnetic radiation are proposed.**

## I. Introduction

Crystalline electro-optic cells are the basic components of modulators of high power optical radiation used in systems of guidance and communications. High efficiency of modulation of these devices is mainly ensured by using of new high-performance electro-optic crystals. However, in some cases the high efficiency can be achieved due to optimization of geometry of electro-optic interaction in well-known materials. Until recently no attention was given to the problems of optimization of elec­to- and acousto-optic cells geometry. For instance, they were not considered even for classic electro-optic materials like lithium niobate. Usually the so called direct cuts of crystals are used in standard devices, at that the samples have the shape of a rectangular prism and their edges are parallel to the crystal-physics axes. However, the analysis (see e.g. [1, 2]) shows that this geometry is not always optimal. The optimized geometry of the cell should ensure maximal intensity of electro-optic interaction and maximal modulation depth that can be achieved by using of non-trivial indirect cuts of crystals. The determination of such cuts sometimes allows to obtain the modulation efficiency several times higher than the one for the cells with direct cuts.

High-efficient electro-optic crystals frequently belong to the low-symmetry classes and are characterized by significant anisotropy [3–5]. Optimal using of these crystals in optoelectronic devices requires detailed experimental studies of manifestations of electro-optic effect (EOE) and further analysis of the spatial distribution of EOE. It can be done by filling of matrices of EOE [6–10], i.e. by experimental determination of all non-zero conponents of a tensor of EOE (electro-optic coefficients (EOC) $r_{unh}$ describing linear EOE) and using of them as a basis for construction of indicative and extreme surfaces of EOE [1, 2, 11]. However, even though the full set of tensor components is known, the optimization of geometry of electro-optic cells is not easy to implement.

## II. Method and object of investigation

The tensors of higher ranks are represented by indicative surfaces (IS) [12]. Mathematically the formula of IS corresponds to the transformation law of tensor components in transition from one coordinate system to another. In the case of the tensor of third rank (EOE) this law is described by the expression:

$$r_{unh} = \alpha_{up}\alpha_{nc}\alpha_{hl}\, r_{pcl} \qquad (1)$$

To construct IS on the basis of (1), its right part have to be expanded, taking into account non-zero and independent EOC $r_{unh}$. At that the direction cosines $\alpha_{up}$, $\alpha_{nc}$, $\alpha_{hl}$ have to be written in spherical coordinate system $(\theta, \varphi)$:

$$\alpha_{r1} = \sin\theta\cos\varphi, \quad \alpha_{r2} = \sin\theta\sin\varphi, \quad \alpha_{r3} = \cos\theta. \qquad (2)$$

Depending on geometry of electro-optic interaction, i.e. mutual orientation of light polarization $\vec{i}$ and electric field $\vec{E}$ (collinear, orthogonal or arbitrary one), the expressions for IS of longitudinal, transversal or arbitrary EOE can be obtained from (1) [1, 13]. They describe the surfaces which are relatively easy constructed in $(\theta, \varphi)$ coordinates using standard (or special) software and completely characterize EOE in crystals of arbitrary symmetry [1].

Uniaxial crystals of lead germanate ($Pb_5Ge_3O_{11}$) belong to the trigonal crystal system (point group 3) [14]. These crystals are interesting for acousto-optic applications, because of the high values of refractive indices $n_1 = 2.11$ and $n_3 = 2.15$ that correspond to the high enough acousto-optic figure-of-merit $M_2$. The thermal expansion coefficients of these crystals are relatively small [14], so the effects of static photoelasticity relaxation are practically absent even at the temperatures closed to the phase transition one (~177 ºC) [15] that is important for practical applications. Moreover, these crystals are piezoelectric that allows to realize the acousto-optic cells on own piezoelectric effect. They are mechanically strong and resistant to aggressive media. Their transparency region of 0.5…4 μm makes them competitive for electro-optic modulation in infrared spectral region. Here we analyze anisotropy of EOE in lead germanate crystals using IS's, determine the maximal values of EOE and corresponding geometries of electro-optic interaction for practical realization of the effective crystalline cell of electro-optic modulator.

III. RESULTS

The matrix of electro-optic coefficients for $Pb_5Ge_3O_{11}$ contains 13 components, six of which are independent [14]:

$$[r_{unh}] \equiv [r_{il}] = \begin{bmatrix} r_{11} & -r_{22} & r_{13} \\ -r_{11} & r_{22} & r_{13} \\ 0 & 0 & r_{33} \\ r_{41} & r_{51} & 0 \\ r_{51} & -r_{41} & 0 \\ -r_{22} & -r_{11} & 0 \end{bmatrix} = \begin{bmatrix} 0.27 & -2.3 & 10.5 \\ -0.27 & 2.3 & 10.5 \\ 0 & 0 & 15.3 \\ 0.6 & 6.0 & 0 \\ 6.0 & -0.6 & 0 \\ -2.3 & -0.27 & 0 \end{bmatrix} \text{pm/V}. \quad (3)$$

If the directions of electrical field $\vec{E}$ and polarization $\vec{i}$ are determined by the angles $(\theta_l, \varphi_l)$ and $(\theta_i, \varphi_i)$ correspondingly, the expression for IS can be obtained from (1), taking into account the independent components of EOE matrix for symmetry class 3. The lengths of radius-vector $r'$ of this surface correspond to the values of EOE for each direction of polarization under the action of electric field in any given one:

$$r'(\theta_i, \varphi_i, \theta_l, \varphi_l) = \sin^2\theta_i [(r_{11}\sin\theta_l\cos\varphi_l - r_{22}\sin\theta_l\sin\varphi_l + r_{13}\cos\theta_l)\cos^2\varphi_i + (-r_{11}\sin\theta_l\cos\varphi_l + r_{22}\sin\theta_l\sin\varphi_l + r_{13}\cos\theta_l)\sin^2\varphi_i + (-r_{11}\sin\varphi_l - r_{22}\cos\varphi_l)\sin\theta_l\sin2\varphi_i] + r_{33}\cos\theta_l\cos^2\theta_i + [(r_{41}\cos\varphi_l + r_{51}\sin\varphi_l)\sin\theta_l\sin\varphi_i + (r_{51}\cos\varphi_l - r_{41}\sin\varphi_l)\sin\theta_l\cos\varphi_i]\sin2\theta_i. \quad (4)$$

If the electric field $\vec{E}$ is applied along $X_1$ axis ($\theta_l = 90°$, $\varphi_l = 0$), the general IS of EOE is described by the expression:

$$r'(\theta_i, \varphi_i, 90°, 0) = \sin^2\theta_i (r_{11}\cos2\varphi_i - r_{22}\sin2\varphi_i) + \sin2\theta_i (r_{41}\sin\varphi_i + r_{51}\cos\varphi_i). \quad (5)$$

This surface (Fig. 1,a) is significantly anisotropic and the maxima of EOE lie in the plane rotated around the optical axis on the small angle of about 6° determined by $r_{61} = r_{22}$ coefficient and simultaneously tilted to the optical axis on the angle of about 22° determined by $r_{41}$. In this case the maxima of EOE are firstly determined by $r_{51}$ coefficient and commensurate with its value, $r_{il}'^{max}(47°, -6°) = 6.29$ pm/V.

If the direction of $\vec{E}$ coincides with $X_1$ axis ($\theta_l = 90°$, $\varphi_l = 90°$), the general EOE IS is described by:

$$r'(\theta_i, \varphi_i, 90°, 90°) = -\sin^2\theta_i (r_{22}\cos2\varphi_i + r_{11}\sin2\varphi_i) + \sin2\theta_i (r_{51}\sin\varphi_i - r_{41}\cos\varphi_i). \quad (6)$$

This surface is shown in Fig. 1,b. Like the previous one, it is considerably anisotropic. The extrema of EOE do not lie in the main planes, e.g. the maximum of 7.3 pm/V is observed in the direction determined by spherical angles $\theta_i = 51°$, $\varphi_i = 96°$.

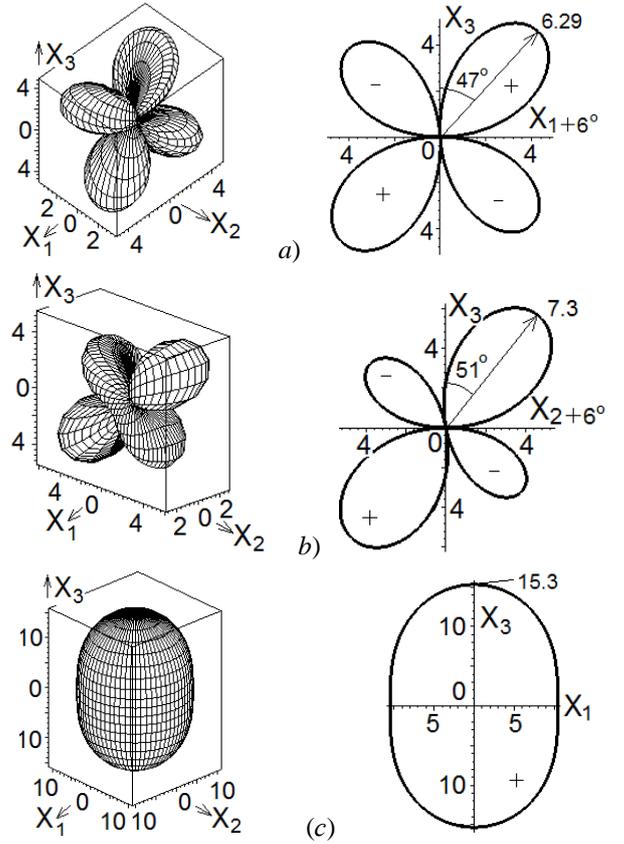

Fig. 1. EOE $r'_{il}$ surfaces of $Pb_5Ge_3O_{11}$ crystals (pm/V) and their cross-sections by the planes with maximal EOE values; the directions of electric field are: (a) $\vec{E} \parallel X_1$ ($\theta_l = 90°$, $\varphi_l = 0$), (b) $\vec{E} \parallel X_2$ ($\theta_l = 90°$, $\varphi_l = 90°$), (c) $\vec{E} \parallel X_3$ ($\theta_l = 0$).

If the electric field $\vec{E}$ is applied along $X_3$ axis ($\theta_l = 0$), the general EOE surface is described by the expression:

$$r'(\theta_i, \varphi_i, 0, 0) = r_{13}\sin^2\theta_i + r_{33}\cos^2\theta_i. \quad (7)$$

The surface constructed on the basis of $r_{il}$ EOCs (3) in accordance with (7) is shown in Fig. 1,c. The maximal value of EOE is equal to 15.3 pm/V, and the corresponding radius-vector is collinear to $X_3$ axis.

From the practical point of view, the collinear or orthogonal mutual orientations of the electric field and the light polarization are particularly important. The expressions for IS's of longitudinal and the transversal EOE in crystals of symmetry class 3 can be easily obtained from (2).

Particularly, substituting $\theta_l = \theta_i$, $\varphi_l = \varphi_i$ in (4), we obtain the formula for collinear EOE conditions when the directions of electric field $\vec{E}$ and the polarization $\vec{i}$ coincide:

$$r'(\theta_i, \varphi_i) = [r_{11}(\cos^3\varphi_i - 3\sin^2\varphi_i\cos\varphi_i) - r_{22}(3\cos^2\varphi_i\sin\varphi_i - \sin^3\varphi_i)]\sin^3\theta_i + r_{13}\sin^2\theta_i\cos\theta_i + r_{33}\cos^3\theta_i + r_{51}\sin\theta_i\sin2\theta_i. \quad (8)$$

The surface of collinear EOE and its cross-section are shown in Fig. 2,a.

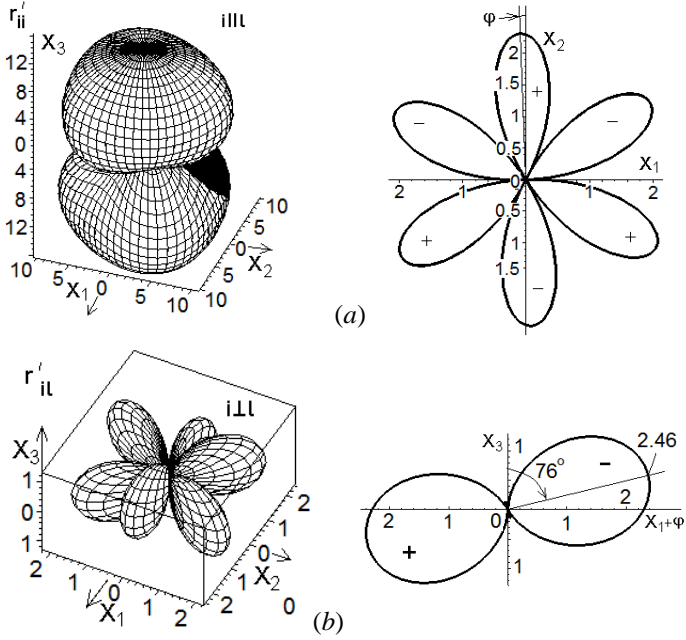

Fig. 2. IS of the collinear EOE $r'_{ii}$ and its cross-section by $X_1, X_2$ plane (*a*); IS of polarization of transversal EOE $r'_{il}$ and its cross-section by $X_1, X_3$ plane (*b*), all values on axes in pm/V, the angle $\varphi = 2°$.

For transversal EOE, the expressions can be written for few types of IS's determined by mutual orientations of the vectors $\vec{i}$ and $\vec{E}$. Particularly, if they are determined by the angles ($\theta_l = 90°$, $\varphi_l = 0…360°$) and ($\theta_i = 0…180°$, $\varphi_i = \varphi_l - 90°$), $\vec{E} \perp \vec{i}$, the following expression for IS is obtained:

$$r'(\theta_i, \varphi_i, 90°, \varphi_i + 90°) = [r_{11}(\sin^3 \varphi_i - 3\sin \varphi_i \cos^2 \varphi_i) + \\ + r_{22}(3\sin^2 \varphi_i \cos \varphi_i - \cos^3 \varphi_i)]\sin^2 \theta_i - r_{41}\sin 2\theta_i. \quad (9)$$

In the case of transversal EOE, this surface is called the surface of light polarization; it is shown in Fig.2,*b*. In the same way the surface of the electric field for the transversal EOE can be constructed, in this case $\vec{E}$ ($\theta_l = 0…360°$, $\varphi_l = 0…360°$) and $\vec{i}$ ($\theta_i = 90°$, $\varphi_i = \varphi_l + 90°$). Substituting these conditions to (2), we obtain the formula for IS of electric field $r'^E_{il}$:

$$r'(90°, \varphi_l + 90°, \theta_l, \varphi_l) = [r_{11}(3\sin^2 \varphi_l \cos \varphi_l - \cos^3 \varphi_l) + \\ - r_{22}(\sin^3 \varphi_l - 3\cos^3 \varphi_l \sin \varphi_l)]\sin^2 \theta_l + r_{13}\sin \theta_l. \quad (10)$$

This surface is shown in Fig. 3.

Based on the formula (2) one can obtain IS under the condition $\vec{E} \perp \vec{i}$, substituting the angles $\theta_l = \theta_i + 90°$, $\varphi_l = \varphi_i$. Obviously, in this case, the direction of electric field coincides with the one of light wave vector. The expression for this surface contains all EOCs of the crystal, because under these conditions the vectors $\vec{E}$ and $\vec{i}$ pass throw all possible directions, but remain orthogonal. This expression is

$$r'(\theta_i, \varphi_i, \theta_i, \varphi_i) = [r_{11}(\cos^3 \varphi_i - 3\sin^2 \varphi_i \cos \varphi_i) - \\ - r_{22}(3\cos^2 \varphi_i \sin \varphi_i - \sin^3 \varphi_i)]\sin^2 \theta_i \cos \theta_i - \\ - r_{13}\sin^3 \theta_i + r_{33}\sin \theta_i \cos^2 \theta_i + r_{51}\cos \theta_i \sin 2\theta_i. \quad (11)$$

and the corresponding surface is shown in Fig. 4.

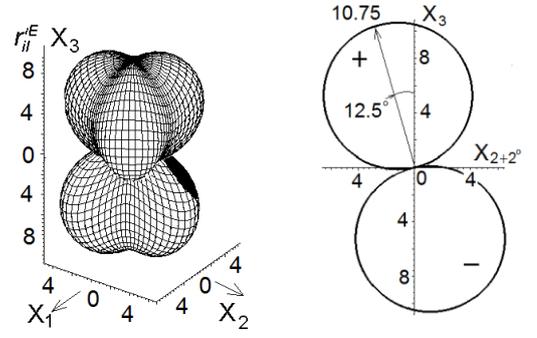

Fig. 3. IS of the electric field for transversal EOE $r'^E_{il}$ and its cross-section by $X_{2+2°}, X_3$ plane (all values on axes in pm/V).

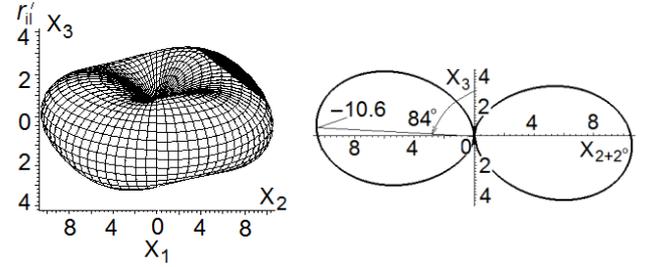

Fig. 4. IS of transversal EOE constructed in accordance with (11) and its cross-section by $X_{2+2°}, X_3$ plane (all values on axes in pm/V).

In practice this geometry of interaction can be realized by using of transparent electrodes ensuring the coincidence of the directions of light propagation and electrical field.

## IV. DISCUSSION

Having analyzed the IS of EOE in lead germanate crystals, the following conclusions can be formulated.

1. IS's described by equations (4) – (8) are considerable anisotropic. These IS's are visually similar to the ones of lithium niobate [13], however, the determination of extreme values for $Pb_5Ge_3O_{11}$ is essentially complicated because of the presence of rotation coefficient $r_{61} = r_{22}$ in its matrix of EOC. Indeed, whereas the analysis of EOE in widely used lithium niobate crystals consists in consideration of surfaces cuts by main planes, obtaining of corresponding expressions and determination of maximal values of the effect by their investigation on extreme points, the analogous analysis of IS's of lead germanate requires using of numerical methods [16] and special software [11]. As it is shown by this analysis, the maximal values of EOE in $Pb_5Ge_3O_{11}$ crystals are generally observed under nonorthogonal conditions of electro-optic interaction. Moreover, these maxima belong to the planes tilted in relation to the crystal-physics axes and the direction of electrical field. The only exception is the surface described by equation (7), see Fig. 1,*c*. However, it is in this geometry, when the directions of electric field and light polarization coincide with $X_3$ axis, the highest electrically induced change of refractive index (corresponds to 15.3 pm/V) is obtained.

2. As it is seen from Figs. 2–4, the indicative surfaces of EOE in lead germanate for orthogonal experimental conditions (8) – (11) differ from the ones of lithium niobate [1] which also belongs to trigonal crystal system. Indeed, the symmetry of $Pb_5Ge_3O_{11}$ (symmetry class 3) is lower than the one of $LiNbO_3$ (3m), so IS's of EOE in lead germanate reveal peculiarities caused by lowering of symmetry. In particular, they are the following: a) the symmetry planes of IS's of longitudinal and transversal EOE do not coincide with the main crystal-physics planes of the crystal, they are rotated around $X_3$ axis on the angle of 2°; b) in the case of transversal EOE in $Pb_5Ge_3O_{11}$ (Fig. 2,b) the maxima of the effect are not in $X_1,X_2$ plane, like in $LiNbO_3$, and the petals of IS with different signs are tilted from this plane on the angle of 14° in opposite directions.

3. The maximal values of longitudinal and transversal EOE in $Pb_5Ge_3O_{11}$ are determined numerically, they are equal to 15.3, 2.46, 10.75 and –10.6 pm/V correspondingly (see Figs. 2–4). The position of maximum coincides with crystal-physics axis ($X_3$ one) only for longitudinal EOE (Fig. 2,a), whereas in the other cases the maxima are observed in non-trivial directions which are not coincide with the crystal-physics axes or the main planes of the crystal (see Figs.2,b, 3, 4).

4. In the case of collinear electro-optic interaction, when the directions of electric field and light propagation coincide (see Fig. 4), IS has got a peculiarity which consists in constant value of EOE in the main $X_1,X_2$ plane, contrary to the previous cases where the effect in $X_1,X_2$ plane is essentially anisotropic (see Fig. 2, 3). The formula for the cross-section of corresponding IS by $X_1,X_2$ plane is obtained by substitution of the value of $\theta_i = 90°$ to (11):

$$r'(90°, \varphi_i, 0°, \varphi_i) = -r_{13}\cos^2\varphi_i - r_{13}\sin^2\varphi_i = -r_{13} = -10.5. \quad (12)$$

As it is seen, this cross-section is a circle with a radius corresponding to $r_{13}$ coefficient. This geometry does not require precise orientation of the sample in the production of electro-optic cells and, taking into account the high value of EOE $r' = -10.5$ pm/V, can be considered as a convenient for practical using.

## V. CONCLUSIONS

The obtained results lead to the following conclusions: lead germanate crystals reveal the value of EOE twice lower than the one in the widely used lithium niobate crystal [1, 14], but is in order higher than the value of EOE in optical-damage-resistant lithium tetraborate crystals [9, 17–20]. The simple geometry of effective working elements (direct cuts for interferometric schemes [21–23], despite the considerable anisotropy of optical properties), high optical quality, manufacturability and wide transparency range indicate the promise of these crystals using in electro-optic devices for infrared spectral region.

**Acknowledgment.** This research has received funding from the European Union's Horizon 2020 research and innovation programme under the Marie Skłodowska-Curie grant agreement No 778156 and from Ministry of Education and Science of Ukraine in the frames of projects "Modulator" (0117U004456) and "Nanocrystalit" (0119U002255).

## REFERENCES

[1] A.S. Andrushchak, B.G. Mytsyk, N.M. Demyanyshyn, M.V. Kaidan, O.V. Yurkevych, S.S. Dumych, A.V. Kityk, and W. Schranz "Spatial anisotropy of linear electro-optic effect for crystal materials: II. Indicative surfaces as efficient tool for electro-optic coupling optimization," Optics & Lasers in Engineering, vol. 47, pp. 24-30, 2009.

[2] N.M. Demyanyshyn, B.G. Mytsyk, Ya.P. Kost', I.M. Solskii, and O.M. Sakharuk, "Elasto-optic effect anisotropy in calcium tungstate crystals," Appl. Opt., vol. 54, pp. 2347-2355, 2015.

[3] D.E. Gray (Ed.), AIP Handbook of Physics, New York: McGraw Hill, 1972.

[4] W.R. Cook, R.F.S.Hearmon, H. Jaffe, and D.F. Nelson, "Piezooptic and electrooptic coefficient constants ", Landolt-Börstein, Group III, Vol. 11, Hellewege, K.-H. and Hellewege, A. M., Eds. (Springer-Verlag, New York, 1979), p. 495.

[5] M. Bass, Handbook of Optics, 2nd ed. McGraw-Hill, 1995.

[6] T.S. Narasimhamurty, Photoelastic and Electro-Optic Properties of Crystals. New York: Plenum Press, 1981.

[7] A.S Sonin, and A.S. Vasilevskaya, Electro-optic crystals. Moscow: Atomizdat, 1971 (in Russian).

[8] A.S. Andrushchak, B.G. Mytsyk, N.M. Demyanyshyn, M.V. Kaidan, O.V. Yurkevych, I.M. Solskii, A.V. Kityk, and W. Schranz, "Spatial anisotropy of linear electro-optic effect for crystal materials: I. Experimental determination of electro-optic tensor by means of interferometric technique," Optics & Lasers in Engineering, vol. 47, pp. 31-38, 2009.

[9] T. Shiosaki, M. Adachi, H. Kobayashi, K. Araki, and A. Kawabata, "Elastic, piezoelectric, acousto-optic and electro-optic properties of $Li_2B_4O_7$," Jap. J. Appl. Phys., vol. 24, pp. 25-27, 1985.

[10] B.G. Mytsyk, Y.P. Kost', N.M. Demyanyshyn, A.S. Andrushchak, and I.M. Solskii, "Piezo-optic coefficients of $CaWO_4$ crystals," Crystallogr. Rep., vol. 60, pp. 130-137, 2015.

[11] O. Buryy, N. Demyanyshyn, B. Mytsyk, and A. Andrushchak, "Optimizing of the piezo-optic interaction geometry in $SrB_4O_7$ crystals," Optica Applicata, vol. 46, pp. 447-459, 2016.

[12] L.A. Shuvalov, A.A. Urusovskaja, and I.S. Zheludev (Eds.), Modern crystallography, vol. 4. Moscow: Nauka, 1981 (in Russian).

[13] N.M. Demyanyshyn, B.G. Mytsyk, A.S. Andrushchak, and O.V. Yurkevych, "Anisotropy of the electro-optic effect in magnesium-doped $LiNbO_3$ crystals," Crystallogr. Rep., vol. 54, pp. 306-312, 2009.

[14] M.P. Shaskolskaya (Ed.), Acoustic Crystals. Handbook. Moscow: Nauka, p. 506, 1982 (in Russian).

[15] N.M. Demyanyshyn, "Relaxation of static photoelasticity in lead germanate crystals," Ukr. J. Phys., vol. 60, pp. 273-276, 2015.

[16] W.H. Press, B.P. Flannery, S.A. Teukolsky, and W.T. Vetterling, Numerical recipes in Pascal. Cambridge: Cambridge University Press, 2007.

[17] R.Vlokh, Ya. Dyachok, O.Krupych, Ya. Burak, I. Martynyuk-Lototska, A.Andrushchak, and V.Adamiv " Study of laser-induced damage of borate crystals", Ukr. J. Phys.Opt., vol. 4, pp.101-104, 2003.

[18] D. Kasprowicz, J. Kroupa, A. Majchrowski, E. Michalski, M. Drozdowski, and J. Żmija "Elastic and nonlinear optical properties of lithium tetraborate" Cryst. Res. Technol., vol. 38, pp. 374-378, 2003.

[19] T. Sugawara, R. Komatsu, and S. Uda Linear and nonlinear optical properties of lithium tetraborate" Solid State Comm., vol. 107, pp. 233-237, 1998.

[20] K. Otsuka, M. Funami, M. Ito, H. Katsuda, M. Tacano, M. Adachi, and A. Kawabata "Design and evaluation of $Li_2B_4O_7$ surface acoustic wave filter" Jpn. J. Appl. Phys., vol. 34, pp. 2646-2649, 1995.

[21] N.M. Demyanyshyn, B.G. Mytsyk, and O.M. Sakharuk "Elasto-optic effect anisotropy in strontium borate crystals ", Applied Optics, vol. 8, pp. 1620-1628, 2014.

[22] B.G. Mytsyk, A.S. Andrushchak, and Ya.P Kost' "Static photoelasticity of gallium phosphide crystals" Crystallography Reports, vol. 57, pp. 124-130, 2012.

[23] B.G. Mytsyk, Ya.P. Kost', N.M. Demyanyshyn, V.M. Gaba, and O.M. Sakharuk "Study of piezo-optic effect of calcium tungstate crystals the conoscopic method", Optical Materials, vol. 39, pp. 69-73, 2015.